\newif\ifpdf
\ifx\pdfoutput\undefined
\pdffalse % we are not running PDFLaTeX
\else
\pdfoutput=1 % we are running PDFLaTeX
\pdftrue \fi

\newif\iffinal
\finaltrue
\documentclass[reqno,twoside,final]{myjstatphys}
\iffinal\else\usepackage[notref,notcite]{showkeys}\fi
\usepackage{cite}
\usepackage{amsmath}
\usepackage{amsfonts}
\usepackage{amssymb}
\usepackage{graphicx}
\usepackage{graphics}
\usepackage{verbatim}

\IfFileExists{epsf.def}{\input epsf.def}{\usepackage{epsf}}

\IfFileExists{myowntimes.sty}
	{\usepackage{myowntimes}}
	{\usepackage{times}
		\IfFileExists{mathrsfs.sty}
			{\usepackage{mathrsfs}}
			{\newcommand{\mathscr}{\mathcal}}}
\newenvironment{proofsect}[1]
{\par\normalfont\vskip0.3cm\noindent
{\hskip10.4mm\sffamily\slshape#1.}}
{\qed\vspace{0.15cm}}

\theoremstyle{jsp}
\newtheorem{proposition}[theorem]{Proposition}

\newcommand{\textd}{\text{\rm d}}

\renewcommand{\AA}{\mathcal A}
\newcommand{\BB}{\mathcal B}
\newcommand{\CC}{\mathcal C}

\newcommand{\EE}{\mathcal E}

\newcommand{\MM}{\mathcal M}

\newcommand{\OO}{\mathcal O}

\newcommand{\VV}{\mathcal V}

\newcommand{\E}{\mathbb E}

\newcommand{\BbbP}{\mathbb P}

\newcommand{\R}{\mathbb R}

\newcommand{\Z}{\mathbb Z}

\newcommand{\scrF}{\mathscr{F}}
\newcommand{\scrG}{\mathscr{G}}

\newcommand{\scrS}{\mathscr{S}}

\newcommand{\twoeqref}[2]{(\ref{#1}--\ref{#2})}
\newcommand{\1}{{\text{\sf 1}}}
\newcommand{\sI}{{\small\textsl{I}}}
\newcommand{\sL}{{\small\textsl{L}}}
\newcommand{\sS}{{\small\textsl{S}}}
\newcommand{\barsS}{\,\overline{\!\small\textsl{S}}}

\newcommand{\mysb}[2] {\mathchoice%
	{#1_{\text{\fontsize{6}{6}\selectfont\rm #2}}}
	{#1_{\text{\fontsize{6}{6}\selectfont\rm #2}}}
	{#1_{\text{\fontsize{5}{6}\selectfont\rm #2}}}
	{#1_{\fontsize{5}{6}\selectfont\text{\rm #2}}} }
\newcommand{\muL}{\mysb{\mu}{L}}
\newcommand{\muS}{\mysb{\mu}{S}}
\newcommand{\alphaI}{\mysb{\alpha}{I}}
\newcommand{\alphaL}{\mysb{\alpha}{L}}

\newcommand{\Jc}{J_{\text{\rm c}}}

\newcommand{\mstar}{m_\star}
\newcommand{\frakmstar}{\mathfrak{m}\mkern1mu}
\newcommand{\frakhstar}{\mathfrak{h}\mkern1mu}

\newcommand{\frakZ}{\mathfrak{Z}}

\newcommand{\ccdot}{\mkern1mu\cdot\mkern1.3mu}
\newcommand{\cc}{{\text{\rm c}}}

\begin{document}

\title[Colligative properties of solutions]{Colligative properties of solutions:\\*[2mm]I.~Fixed concentrations}

\author{Kenneth~S.~Alexander,\footnotemark[1] Marek~Biskup,\footnotemark[2] and Lincoln~Chayes\footnotemark[2]}

\renewcommand{\thefootnote}{}
\footnotetext{\hglue-1.9em$\copyright$\,2004 by K.S.~Alexander, M.~Biskup and
L.~Chayes.  Reproduction, by any means, of the entire article for
non-commercial purposes is permitted  without~charge.}
\renewcommand{\thefootnote}{\arabic{footnote}}

\footnotetext[1]{Department of Mathematics, USC, Los  Angeles, California, USA}
\footnotetext[2]{Department of Mathematics, UCLA, Los Angeles, California, USA.}

\runningauthor{Alexander, Biskup and Chayes}

\begin{abstract}
Using the formalism of rigorous  statistical
mechanics, we study the phenomena  of phase separation and freezing-point
depression upon  freezing of solutions. Specifically, we devise an
Ising-based model of a solvent-solute  system and show  that, in the
ensemble with a fixed amount of solute, a macroscopic phase  separation
occurs in an interval of values of the chemical potential of the solvent.
The boundaries of  the phase separation domain in the phase diagram are
characterized and shown to asymptotically  agree with the formulas used in
heuristic analyses of  freezing point depression. The limit of
infinitesimal concentrations is described in a subsequent paper.
\end{abstract}

\section{Introduction}
\subsection{Motivation}
\label{sec1.1}\noindent 
The statistical mechanics of pure  systems---most
prominently the topic of phase  transitions and their associated surface
phenomena---has been  a subject of  fairly intensive research in recent
years. Several physical principles for  pure systems (the Gibbs phase rule,
Wulff construction, etc.)  have been put on a mathematically  rigorous
footing and, if necessary,  supplemented with appropriate conditions
ensuring their  validity. The corresponding phenomena in systems with
several mixed components,  particularly  solutions, have long been
well-understood   on the level of theoretical physics. However, they  have
not received much mathematically rigorous attention and  in particular have
not been  derived   rigorously starting from a local interaction. A natural
task is to use the ideas from  statistical mechanics of pure systems to
develop a higher level of control for phase transitions in  solutions. This
is especially desirable in light of the important role that basic  physics
of these systems plays in sciences, both general (chemistry, biology,
oceanography) and applied  (metallurgy, etc.). See e.g.
\cite{Moore,Landau-Lifshitz,Cottrell} for more discussion.

Among the perhaps most interesting aspects of phase transitions in  mixed
systems is a dramatic
\emph{phase separation} in solutions upon freezing (or boiling). A
well-known example from ``real world'' is the formation of brine pockets in
frozen sea water. Here,  two important physical phenomena are observed:
\begin{enumerate}
\item[(1)] 
Migration of nearly all the salt into whatever portion of
ice/water mixture remains liquid.
\item[(2)] 
Clear evidence of \emph{facetting} at the water-ice boundaries.
\end{enumerate}
Quantitative analysis also reveals the following fact:
\begin{enumerate}
\item[(3)] 
Salted water freezes at temperatures lower than the  freezing
point of pure water. This is the phenomenon of \emph{freezing point
depression}.
\end{enumerate}
Phenomenon~(1) is what ``drives'' the physics of sea ice and is thus
largely responsible for the variety of physical effects
that have been observed, see e.g.~\cite{Golden,GoldenAckleyLytle}.
Notwithstanding, (1--3) are not special to the salt-water system;
they are shared by a large class of the so called \emph{non-volatile}
solutions. A  discussion concerning the general aspects of freezing/boiling
of solutions---often  referred to as
\emph{colligative} properties---can be found
in~\cite{Moore,Landau-Lifshitz}.

Of course, on a heuristic level, the above phenomena are far from
mysterious. Indeed, (1) follows from the observation that, macroscopically,
the liquid phase  provides a more hospitable environment for salt than the
solid phase. Then (3) results by noting  that the migration of salt
increases the entropic  cost of freezing so the energy-entropy  balance
forces the transition point  to a lower temperature.  Finally, concerning
observation~(2)  we note that, due to the crystalline  nature of ice, the
ice-water surface tension will be  anisotropic. Therefore, to describe the
shape of brine pockets, a Wulff construction  has to be  involved with the
caveat that here the crystalline phase is on the outside. In summary, what
is underlying these phenomena is a phase separation  accompanied by the
emergence of a crystal  shape. In the context of pure systems, such topics
have been well understood at the  level of  theoretical physics for quite
some time \cite{Wulff,Curie,Gibbs,Wortis} and, recently (as measured  on the
above time scale), also at the level of rigorous theorems in
two~\cite{ACC,DKS,Pfister,Pf-Velenik,Bob+Tim,bigBCK} and
higher~\cite{Cerf,Bodineau,Cerf-Pisztora} dimensions.

The purpose of this and a subsequent paper is
to study the qualitative nature of phenomena (1--3) using the formalism of equilibrium
statistical mechanics. 
Unfortunately, a  microscopically realistic model of
salted water/ice system is far beyond reach of rigorous  methods.  (In fact,
even in pure water, the phenomenon of freezing is so complex that crystalization
in realistic models has only recently---and only marginally---been exhibited in 
computer simulations~\cite{MatsumotoSaitoOhmine}.)  
Thus we will resort to a simplified
version in which salt and both phases of water are represented by discrete
random variables residing at sites of a regular lattice. For these models
we show that phase  separation dominates a non-trivial
\emph{region} of chemical potentials in the phase diagram---a  situation
quite unlike the pure system where phase separation can occur only at a
single value  (namely, the transition value) of the chemical potential. The
boundary lines of the phase-separation  region can be explicitly
characterized and shown to agree with the approximate solutions of  the
corresponding problem in the physical-chemistry literature.

The above constitutes the subject of the present paper. In a  subsequent
paper \cite{ABC2} we will demonstrate that, for infinitesimal salt
concentrations scaling  appropriately with the size of the system, phase
separation may still occur dramatically in the  sense that a non-trivial
fraction of the system suddenly melts (freezes) to form a pocket (crystal). In these
circumstances the amount of salt needed is proportional to the
\emph{boundary} of the system which  shows that the onset of freezing-point
depression is actually a surface phenomenon. On a  qualitative level, most
of the aforementioned conclusions should apply to general non-volatile
solutions under the conditions when the solvent freezes (or boils).
Notwithstanding, throughout this  and the subsequent paper we will adopt the
\emph{language} of salted water and refer to the solid  phase of the solvent
as ice, to the liquid phase as liquid-water, and to the solute as salt.

\subsection{General Hamiltonian}
\label{sec1.2}\noindent Our model will be defined on the
$d$-dimensional hypercubic lattice~$\Z^d$. We will take the (formal)
nearest-neighbor  Hamiltonian of the following form:
\begin{equation}
\label{Hgen}
\beta\mathscr{H} =-\sum_{\langle x,y\rangle} (\alphaI\sI_x\sI_y
+\alphaL\sL_x\sL_y ) +\kappa\sum_x\sS_x\sI_x
-\sum_x\muS\sS_x-\sum_x\muL\sL_x.
\end{equation}
Here $\beta$ is the inverse temperature (henceforth
incorporated into the Hamitonian),~$x$ and~$y$ are sites in~$\Z^d$
and~$\langle x,y\rangle$  denotes a neighboring pair of sites. The
quantities~$\sI_x$,~$\sL_x$ and~$\sS_x$ are the ice  (water), liquid (water)
and salt variables, which will take values in~$\{0,1\}$ with the  additional
constraint
\begin{equation}
\label{L+I}
\sI_x+\sL_x=1
\end{equation}
valid at each site~$x$. We will say that~$\sI_x=1$  indicates
the \emph{presence of ice} at~$x$ and,  similarly,~$\sL_x$ the
\emph{presence of liquid}  at~$x$. Since a single water molecule cannot
physically be in an ice state, it is  natural to  interpret the
phrase~$\sI_x=1$ as referring to the collective behavior of many particles
in the  vicinity of~$x$ which are enacting an ice-like state, though  we do
not formally incorporate  such a viewpoint into our model.

The various terms in \eqref{Hgen} are essentially self-explanatory:  An
interaction between neighboring ice points, similarly for neighboring liquid
points (we  may assume these to be attractive), an energy penalty~$\kappa$
for a simultaneous presence  of salt and ice at one point, and, finally,
fugacity terms for salt and liquid. For simplicity (and  tractability),
there is no direct salt-salt interaction, except for the exclusion rule of
at  most one salt ``particle'' at each site. Additional terms which could
have been included are  superfluous due to the constraint
\eqref{L+I}. We will assume throughout that~$\kappa>0$, so that the
salt-ice interaction expresses the negative affinity of salt to the ice
state of water.  This term is entirely---and not subtly---responsible for
the general phenomenon of freezing point  depression. We remark that by
suitably renaming the variables, the Hamiltonian in \eqref{Hgen}  would just
as well describe a system with boiling point elevation.

As we said, the variables~$\sI_x$ and~$\sL_x$ indicate the presence  of ice
and liquid water at site~$x$, respectively. The assumption~$\sI_x+\sL_x=1$
guarantees  that \emph{something} has to be present at~$x$ (the
concentration of water in water is unity); what  is perhaps unrealistic is
the restriction of~$\sI_x$ and~$\sL_x$ to only the extreme values,
namely~$\sI_x,\sL_x\in\{0,1\}$. Suffice it to say that the authors are
confident (e.g., on the basis  of \cite{BCG}) that virtually all the results
in this note can be extended to the cases  of continuous variables. However,
we will not make any such mathematical claims; much of this  paper will rely
heavily on preexisting technology which, strictly speaking, has only been
made  to work for the discrete case. A similar discussion applies, of
course, to the salt variables.  But here our restriction
to~$\sS_x\in\{0,1\}$ is mostly to ease the exposition; virtually all  of our
results directly extend to the cases when~$\sS_x$ takes arbitrary (positive)
real  values according to some \emph{a priori} distribution.

\subsection{Reduction to Ising variables}
\label{sec1.3}\noindent It is not difficult to see that the  ``ice-liquid
sector'' of the general Hamiltonian \eqref{Hgen} reduces to a ferromagnetic
Ising spin  system. On a formal level, this is achieved by passing to the Ising variables
$\sigma_x=\sL_x-\sI_x$, which in light of  the constraint \eqref{L+I} gives
\begin{equation}
%\label{}
\sL_x=\frac{1+\sigma_x}2
\quad\text{and}\quad
\sI_x=\frac{1-\sigma_x}2.
\end{equation}
By substituting these into \eqref{Hgen}, we arrive at  the
interaction Hamiltonian:
\begin{equation}
\label{1.4}
\beta\mathscr{H}= -J\sum_{\langle x,y\rangle}\sigma_x\sigma_y
-h\sum_x\sigma_x +\kappa\sum_x\sS_x\frac{1-\sigma_x}2 -\sum_x\muS\sS_x,
\end{equation}
where the new parameters $J$ and $h$ are given by
\begin{equation}
\label{1.5} J=\frac{\alphaL+\alphaI}4
\quad\text{and}\quad h=\frac d2(\alphaL-\alphaI)+\frac{\muL}2.
\end{equation}
We remark that the third sum in \eqref{1.4} is still  written
in terms of ``ice'' indicators so that~$\mathscr{H}$ will have a well
defined meaning  even if~$\kappa=\infty$, which corresponds to prohibiting
salt entirely at ice-occupied sites.  (Notwithstanding, the bulk of this
paper is restricted to finite~$\kappa$.) Using an appropriate  restriction
to finite volumes, the above Hamitonian allows  us to define the
corresponding Gibbs  measures. We postpone any relevant technicalities to
Section~\ref{sec2.1}.

The Hamiltonian as written foretells the possibility of fluctuations  in the
salt  concentration. However, this is \emph{not} the situation which is of
physical  interest. Indeed, in an open system it is clear that the salt
concentration will, eventually,  adjust itself until the system exhibits a
pure phase. On the level of the description provided by
\eqref{1.4} it is noted that, as grand canonical variables, the salt
particles can be explicitly  integrated, the result being the Ising model at
coupling constant~$J$ and external  field~$h_{\text{eff}}$, where
\begin{equation}
\label{1.6}
h_{\text{eff}}=h+\frac12\log\frac{1+e^{\muS}}{1+e^{\muS-\kappa}}.
\end{equation}
In this context, phase coexistence is confined to the
region~$h_{\text{eff}}=0$, i.e., a simple curve in the $(\muS,h)$-plane.
Unfortunately, as is well 
known~\cite{SS2,GS1,GS2,Kotecky-Medved,BCK-comment}, not much insight  on
the subject of
\emph{phase separation} is to be gained by studying the Ising magnet  in an
external field. Indeed, under (for example) minus boundary conditions,
once~$h$ exceeds a particular value, a droplet will form  which all but
subsumes the allowed volume.  The transitional value of~$h$ scales
inversely with  the linear size of the system; the exact constants and the
subsequent behavior of the  droplet depend on the details of the boundary
conditions.

The described ``failure'' of the grand canonical description  indicates that
the correct ensemble in this case is the one with a fixed amount of salt per
unit volume.  (The technical definition uses conditioning from the grand
canonical measure; see  Section~\ref{sec2.1}.) This ensemble is physically
more relevant because, at the moment of freezing, the salt  typically does
not have enough ``mobility'' to be gradually released from the system. It
is  noted that, once the total amount of salt is fixed, the chemical
potential~$\muS$ drops out of  the problem---the relevant parameter is now
the salt concentration. As will be seen in  Section~\ref{sec2}, in our
Ising-based model of the solvent-solute system, fixing the salt concentration  generically
leads to \emph{sharp} phase separation in the Ising configuration.
Moreover, this  happens for an \emph{interval} of values of the
magnetic field~$h$. Indeed, the interplay between the salt concentration and the
actual external field will demand a particular value of the magnetization, even under
conditions which will force a droplet (or ice crystal, depending on the
boundary condition) into the system.

\begin{remark}
\label{rem0}
We finish by noting that, while the parameter $h$ is formally unrelated to temperature, it does to a limited extent play the role of temperature in that it reflects
the \emph{a  priori} amount of preference of the system for water \emph{vs}
ice.  Thus the natural phase  diagram to study is in the $(c,h)$-plane.
\end{remark}

\subsection{Heuristic derivations and outline}
\label{sec1.4}\noindent The reasoning which led to formula
\eqref{1.6} allows for an immediate heuristic explanation of our principal
results. The key  simplification---which again boils down to the absence  of
salt-salt interaction---is that for any Ising  configuration, the
amalgamated contribution of salt, i.e., the Gibbs weight summed over salt
configurations, depends only on the overall magnetization and not on the
details of how the magnetization  gets distributed about the system. 
In systems of linear
scale~$L$, let~$\frakZ_L(M)$ denote the canonical partition function for the
Ising magnet with  constrained overall magnetization~$M$. The total partition
function~$Z_L(c,h)$ at fixed salt concentration~$c$  can then be written as
\begin{equation}
\label{1.7a} Z_L(c,h)=\sum_{M}\frakZ_L(M)e^{hM}W_L(M,c),
\end{equation}
where~$W_L(M,c)$ denotes the sum of the salt part of  the
Boltzmann weight---which only depends on the Ising spins via the total
magnetization~$M$---over all salt configurations with concentration~$c$.

As usual, the physical values of the magnetization are those bringing  the
dominant contribution to the sum in \eqref{1.7a}. Let us recapitulate the
standard  arguments by first considering the case $c = 0$ (which implies
$W_L = 1$), i.e., the usual Ising system  at external field $h$. Here we
recall that~$\frakZ_L(mL^d)$ can approximately be written as
\begin{equation}
\label{1.8a}
\frakZ_L(mL^d)\approx e^{-L^d[\scrF_J(m)+C]},
\end{equation}
where~$C$ is a suitably chosen constant  and~$\scrF_J(m)$ is
a (normalized) canonical free energy. The principal fact about $\scrF_J(m)$
is that  it vanishes for~$m$ in the interval~$[-\mstar,\mstar]$,
where~$\mstar=\mstar(J)$ denotes the  spontaneous magnetization of the Ising
model at coupling~$J$, while it is strictly positive and  strictly convex
for~$m$ with~$|m|>\mstar$. The presence of the ``flat piece'' on the graph
of~$\scrF_J(m)$ is directly responsible for the existence of the phase
transition in the Ising  model: For~$h>0$ the dominant contribution to the
grand canonical partition function comes from~$M\gtrsim\mstar L^d$ while
for~$h<0$ the dominant values of the overall magnetization
are~$M\lesssim-\mstar L^d$. Thus, once~$\mstar=\mstar(J)>0$---which happens
for~$J>\Jc(d)$  with~$\Jc(d)\in(0,\infty)$ whenever~$d\ge2$---a phase
transition occurs at~$h=0$.

The presence of salt variables drastically changes the entire  picture.
Indeed, as we will see in Theorem \ref{thm1}, the salt partition function
$W_L(M,c)$ will  exhibit a nontrivial exponential behavior which is
characterized by a \emph{strictly convex} free  energy. The resulting
exponential growth rate of $\frakZ_L(M)e^{hM}W_L(M,c)$ for $M\approx
mL^d$ is thus no longer a function with a flat piece---instead, for
each~$h$ there is  a \emph{unique} value of~$m$ that optimizes the
corresponding free energy.  Notwithstanding  (again, due to the absence of
salt-salt interactions) once that~$m$ has been selected, the spin
configurations are the typical Ising configurations with overall
magnetizations~$M\approx mL^d$.  In  particular, whenever~$Z_L(c,h)$ is
dominated by values of~$M\approx mL^d$ for
an~$m\in(-\mstar,\mstar)$, a \emph{macroscopic droplet} develops in  the
system. Thus, due to the one-to-one correspondence between~$h$ and the
optimal value of~$m$,  phase separation occurs for an \emph{interval} of
values of~$h$ at any positive concentration;  see Fig.~\ref{fig1}.

\smallskip 
We finish with an outline of the remainder of this paper  and
some discussion of the companion paper~\cite{ABC2}. In Section~\ref{sec2} we
define  precisely the model of interest and state our main results
concerning the asymptotic behavior of the  corresponding measure on spin and
salt configurations  with fixed concentration of salt. Along with  the
results comes a description of the phase diagram and a discussion of
freezing-point  depression, phase separation, etc., see
Section~\ref{sec2.3}. Our main results are proved in  Section~\ref{sec3}.
In~\cite{ABC2} we investigate the asymptotic of infinitesimal salt
concentrations.  Interestingly, we find that, in order to induce phase
separation, the concentration has to scale at  least as the inverse 
linear size of the system.

\section{Rigorous results}
\label{sec2}
\subsection{The model}
\label{sec2.1}\noindent With the (formal) Hamiltonian \eqref{1.4} in  mind,
we can now start on developing the \emph{mathematical} layout of the
problem. To define  the model, we will need to restrict attention to finite
subsets of the lattice. We will mostly  focus on rectangular boxes
$\Lambda_L\subset\Z^d$ of $L\times L\times\dots\times L$ sites  centered at
the origin. Our convention for the boundary,~$\partial\Lambda$, of the set
$\Lambda\subset\Z^d$ will be the collection of sites outside~$\Lambda$ with
a neighbor  inside~$\Lambda$. For each $x\in\Lambda$, we have the water and
salt variables, $\sigma_x\in\{-1,+1\}$ and
$\sS_x\in\{0,1\}$. On the boundary, we will consider fixed
configurations~$\sigma_{\partial\Lambda}$; most of the time we will be
discussing the cases $\sigma_{\partial\Lambda}=+1$
or~$\sigma_{\partial\Lambda}=-1$, referred to as plus and minus boundary
conditions. Since there is no  salt-salt interaction, we may as well
set~$\sS_x=0$ for all~$x\in\Lambda^\cc$.

We will start by defining the interaction Hamiltonian.
Let~$\Lambda\subset\Z^d$ be a finite set. For a spin
configuration~$\sigma_{\partial\Lambda}$ and the
pair~$(\sigma_\Lambda,\sS_\Lambda)$ of spin and salt configurations, we let
\begin{equation}
\label{2.1a}
\beta\mathscr{H}_\Lambda(\sigma_\Lambda,\sS_\Lambda|\sigma_{\partial\Lambda})=
-J\!\!\!\sum_{\begin{subarray}{c}
\langle x,y\rangle\\ x\in\Lambda,\,y\in\Z^d
\end{subarray}}\!\!
\sigma_x\sigma_y
-h\sum_{x\in\Lambda}\sigma_x+\kappa\sum_{x\in\Lambda}\sS_x\frac{1-\sigma_x}2.
\end{equation}
Here, as before,~$\langle x,y\rangle$ denotes a
nearest-neighbor pair on~$\Z^d$ and the parameters~$J$,~$h$ and~$\kappa$
are as discussed  above. (In light of the discussion from
Section~\ref{sec1.3} the last term in \eqref{1.4} has  been omitted.) The
probability distribution of the pair~$(\sigma_\Lambda,\sS_\Lambda)$  takes
the usual Gibbs-Boltzmann~form:
\begin{equation}
\label{PLambda}
P_\Lambda^{\sigma_{\partial\Lambda}}(\sigma_\Lambda,\sS_\Lambda)
=\frac{e^{-\beta
\mathscr{H}_\Lambda(\sigma_\Lambda,\sS_\Lambda|\sigma_{\partial\Lambda})}}
{Z_\Lambda(\sigma_{\partial\Lambda})},
\end{equation}
where the normalization
constant,~$Z_\Lambda(\sigma_{\partial\Lambda})$, is the partition function.
The distributions in~$\Lambda_L$ with the plus  and minus boundary
conditions will be denoted by~$P_L^+$ and~$P_L^-$, respectively.

For reasons discussed before we will be interested in the problems  with a
fixed salt concentration $c\in[0,1]$. In finite volume, we take this to
mean  that the total amount of salt,
\begin{equation}
\label{1.9} N_L=N_L(\sS)=\sum_{x\in\Lambda_L}\sS_x,
\end{equation}
is fixed. To simplify future discussions, we will  adopt the
convention that ``concentration~$c$'' means that $N_L\le
c|\Lambda_L|<N_L+1$, i.e.,
$N_L=\lfloor cL^d\rfloor$. We may then define the finite volume Gibbs
probability measure with salt  concentration~$c$ and plus (or minus)
boundary conditions denoted by~$P_L^{+,c,h}$  (or~$P_L^{-,c,h}$). In light of
\eqref{PLambda}, these are given by the formulas
\begin{equation}
%\label{}
P_L^{\pm,c,h}(\cdot)=P_L^\pm\bigl(\,\cdot\,\big|N_L=\lfloor
cL^d\rfloor\bigr).
\end{equation}
Both measures~$P_L^{\pm,c,h}$ depend on the  parameters~$J$
and~$\kappa$ in the Hamiltonian. However, we will always regard these as
fixed and  suppress them from the notation whenever possible.

\subsection{Main theorems}
\label{sec2.2}\noindent In order to describe our first set of  results, we
will need to bring to bear a few standard facts about the Ising model. For each spin
configuration
$\sigma=(\sigma_x)\in\{-1,1\}^{\Lambda_L}$ let us define the overall
magnetization in~$\Lambda_L$ by the formula
\begin{equation}
\label{2.1} M_L=M_L(\sigma)=\sum_{x\in\Lambda_L}\sigma_x.
\end{equation}
Let $\frakmstar(h,J)$ denote the magnetization of the  Ising
model with coupling constant~$J$ and external field~$h\ge0$. As is well
known, cf the proof of Theorem~\ref{thm3.1}, $h\mapsto\frakmstar(h,J)$
continuously (and strictly) increases from the  
value of the spontaneous magnetization
$\mstar=\frakmstar(0,J)$ to one as $h$ sweeps through
$[0,\infty)$. In particular, for each $m\in[\frakmstar(0,J),1)$, there
exists a unique
$\frakhstar=\frakhstar(m,J)\in[0,\infty)$ such that
$\frakmstar(\frakhstar,J)=m$.

Next we will use the above quantities to define the
function~$\scrF_J\colon(-1,1)\to[0,\infty)$, which represents the canonical
free energy of the Ising model in~\eqref{1.8a}.  As it turns out---see
Theorem~\ref{thm3.1} in Section~\ref{sec3}---we simply have
\begin{equation}
\label{2.2}
\scrF_J(m)=\int\textd m'\, \frakhstar(m',J)\1_{\{\mstar\le m'\le|m|\}},
\qquad m\in(-1,1).
\end{equation}
As already mentioned, if~$J>\Jc$, where~$\Jc=\Jc(d)$  is the
critical coupling constant of the Ising model, then $\mstar>0$ and thus
$\scrF_J(m)=0$  for $m\in[-\mstar,\mstar]$. (Since~$\Jc(d)<\infty$ only
for~$d\ge2$, the resulting ``flat piece''  on the graph
of~$m\mapsto\scrF_J(m)$ appears only in dimensions~$d\ge2$.) From the
perspective of the large-deviation theory,
cf~\cite{Dembo-Zeitouni,denHollander},
$m\mapsto\scrF_J(m)$ is the large-deviation rate function for the
magnetization in the  (unconstrained) Ising model; see again Theorem~\ref{thm3.1}.

Let $\scrS(p)=p\log p+(1-p)\log(1-p)$ denote the entropy  function
of the Bernoulli distribution with parameter~$p$. (We will set
$\scrS(p)=\infty$  whenever $p\not\in[0,1]$.) For each $m\in(-1,1)$, each
$c\in[0,1]$ and each $\theta\in[0,1]$, let
\begin{equation}
\label{2.3}
\Xi(m,\theta;c)= -\frac{1+m}2\scrS\Bigl(\frac{2\theta c}{1+m}\Bigr)
-\frac{1-m}2\scrS\Bigl(\frac{2(1-\theta)c}{1-m}\Bigr).
\end{equation}
As we will show in Section~\ref{sec3}, this quantity
represents the entropy of configurations with fixed salt concentration~$c$,
fixed overall  magnetization~$m$ and fixed fraction~$\theta$ of the salt
residing ``on the plus spins'' (and  fraction $1-\theta$ ``on the minus
spins'').

\smallskip Having defined all relevant quantities, we are ready to  state
our results. We begin with a large-deviation principle for the magnetization
in the  measures~$P_L^{\pm,c,h}$:

\begin{theorem}
\label{thm1} Let $J>0$ and $\kappa>0$ be fixed. For each $c\in(0,1)$,  each
$h\in\R$ and each
$m\in(-1,1)$, we have
\begin{multline}
\label{2.4}
\quad\lim_{\epsilon\downarrow0}\lim_{L\to\infty}\frac1{L^d}
\log P_L^{\pm,c,h}\bigl(|M_L-mL^d|\le\epsilon L^d\bigr)
\\=
-G_{h,c}(m)+\inf_{m'\in(-1,1)}G_{h,c}(m').
\quad
\end{multline}
Here $m\mapsto G_{h,c}(m)$ is given by
\begin{equation}
\label{Phi} G_{h,c}(m)=\inf_{\theta\in[0,1]}\scrG_{h,c}(m,\theta),
\end{equation}
where
\begin{equation}
\label{Phi2}
\scrG_{h,c}(m,\theta)=-hm-\kappa\theta c-\Xi(m,\theta;c)+\scrF_J(m).
\end{equation}
The function $m\mapsto G_{h,c}(m)$ is finite and  strictly
convex on~$(-1,1)$ with
$\lim_{m\to\pm1}G_{h,c}'(m)=\pm\infty$. Furthermore, the unique  minimizer
$m=m(h,c)$ of $m\mapsto G_{h,c}(m)$ is continuous in both~$c$ and~$h$ and
strictly increasing in~$h$.
\end{theorem}

On the basis of the above large-deviation result, we can now  characterize
the typical configurations of the measures~$P_L^{\pm,c,h}$. Consider the
Ising  model with coupling constant~$J$ and zero external field and
let~$\BbbP_L^{\pm,J}$ be the  corresponding Gibbs measure in
volume~$\Lambda_L$ and~$\pm$-boundary condition. Our main result  in this
section is then as follows:

\begin{theorem}
\label{thm2} Let~$J>0$ and~$\kappa>0$ be fixed. Let $c\in(0,1)$
and $h\in\R$, and define two sequences of probability measures~$\rho_L^\pm$
on~$[-1,1]$ by the formula
\begin{equation}
\label{2.5}
\rho_L^\pm\bigl([-1,m]\bigr)=P_L^{\pm,c,h}(M_L\le mL^d),
\qquad m\in[-1,1].
\end{equation}
The measures $\rho_L^\pm$ allow us to write the spin
marginal of the measure~$P_L^{\pm,c,h}$ as a convex combination of the Ising
measures  with fixed magnetization; i.e., for any set~$\AA$ of
configurations $(\sigma_x)_{x\in\Lambda_L}$, we have
\begin{equation}
\label{2.6}
P_L^{\pm,c,h}\bigl(\AA\times\{0,1\}^{\Lambda_L}\bigr)=\int\!\rho_L^\pm(\textd
m)\,
\BbbP_L^{\pm,J}\bigl(\AA\big|M_L=\lfloor mL^d\rfloor\bigr).
\end{equation}
Moreover, if~$m=m(h,c)$ denotes the unique minimizer  of the
function~$m\mapsto G_{h,c}(m)$ from \eqref{Phi}, then the following properties
are true:
\settowidth{\leftmargini}{(1111)}
\begin{enumerate}
\item[(1)] Given the spin configuration on a finite set
$\Lambda\subset\Z^d$, the salt variables on~$\Lambda$ are
asymptotically independent. Explicitly,  for each finite set
$\Lambda\subset\Z^d$ and any two configurations
$\barsS_\Lambda\in\{0,1\}^\Lambda$  and
$\bar\sigma_\Lambda\in\{-1,1\}^\Lambda$,
\begin{multline}
\label{2.7}
\quad
\lim_{L\to\infty}
P_L^{\pm,c,h}\bigl(\sS_\Lambda=\barsS_\Lambda\big|\sigma_\Lambda=\bar\sigma_\Lambda\bigr)
\\
=\prod_{x\in
\Lambda}\bigl\{q_{\bar\sigma_x}\delta_1(\barsS_x)+(1-q_{\bar\sigma_x})\delta_0(\barsS_x)\bigr\},
\quad
\end{multline}
where the numbers~$q_\pm\in[0,1]$ are uniquely  determined by
the equations
\begin{equation}
\label{2.8}
\frac{q_+}{1-q_+}=\frac{q_-}{1-q_-}\,e^\kappa
\qquad\text{and}\qquad q_+\frac{1+m}2+q_-\frac{1-m}2=c.
\end{equation}
\item[(2)] The measure~$\rho_L^\pm$ converges weakly to a point mass
at~$m=m(h,c)$,
\begin{equation}
\label{2.9}
\lim_{L\to\infty}\rho_L^\pm(\cdot)=\delta_m(\cdot).
\end{equation}
In particular, the Ising-spin marginal of the
measure~$P_L^{\pm,c,h}$ is asymptotically supported  on the usual Ising spin
configurations with  the overall magnetization~$M_L=(m+o(1))L^d$, where~$m$
minimizes~$m\mapsto G_{h,c}(m)$.
\end{enumerate}
\end{theorem}

The fact that conditioning $P_L^{\pm,c,h}$ on a fixed value of
magnetization produces the Ising measure under same conditioning---which is
the content  of~\eqref{2.6}---is directly related to the absence of
salt-salt interaction. The principal conclusions of  the previous theorem
are thus parts~(1) and~(2), which state that the presence of a particular
amount of salt \emph{forces} the Ising sector to choose a particular value
of magnetization density.  The underlying variational principle provides
insight into the physical mechanism of phase  separation upon freezing of
solutions. (We refer the reader back to Section~\ref{sec1.4} for the
physical basis of these considerations.)

\smallskip We will proceed by discussing the consequences of these  results
for the phase diagram of the model and, in particular, the phenomenon of
freezing point  depression. Theorems~\ref{thm1} and~\ref{thm2} are proved in
Section~\ref{sec3.2}.

%%%% COUNTER FOR FIGURES
\newcounter{obrazek}

\begin{figure}[t]
\refstepcounter{obrazek}
\label{fig1}
%\vspace{.2in}
\ifpdf
\centerline{\includegraphics[width=3.5in]{extdiag.pdf}}
\else
\centerline{\includegraphics[width=3.5in]{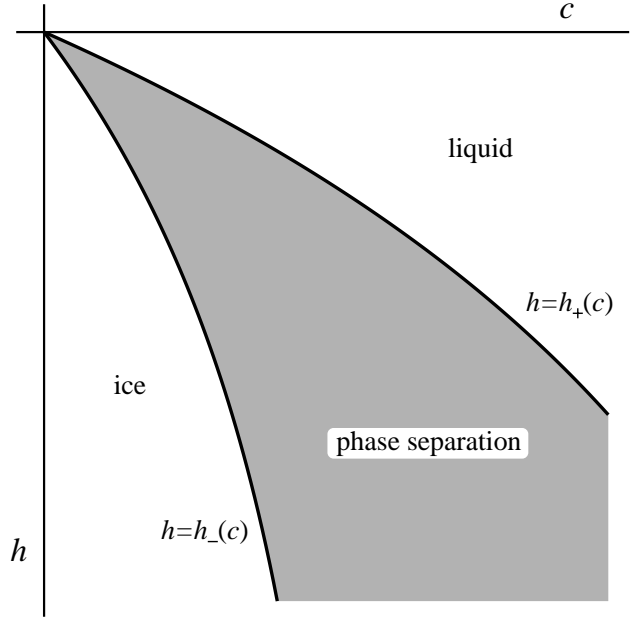}}
\fi
%\vspace{.2in}
\bigskip
%\begin{quote}
%\fontsize{9}{2}\selectfont 
\caption{The phase  diagram of
the ice-water system with $\kappa\gg1$.  The horizontal axis marks the
concentration of  the salt in the system, the vertical line represents the
external field acting on the Ising  spins---see formula~\eqref{1.5}. For
positive concentrations $c>0$, the system stays in the  liquid-water phase
throughout a non-trivial range of negative values of~$h$---a manifestation
of the  freezing-point depression. For~$(h,c)$ in the shaded region, a
non-trivial fraction of the  system is frozen into ice. Once
$(h,c)$ is on the left of the shaded region, the entire system is in  the
ice state. For moderate values of~$\kappa$, the type of convexity of the transition lines
may change from concave to convex near~$(h,c)=(0,0)$; see the companion paper~\cite{ABC2}.
\normalsize
}
\end{figure}

\subsection{Phase diagram}
\label{sec2.3}\noindent The representation \eqref{2.6} along with the
asymptotic \eqref{2.9} allow us to characterize the
distribution~$P_L^{\pm,c,h}$ in terms of  the canonical ensemble of the
Ising ferromagnet. Indeed, these formulas imply that the  distribution of
Ising spins induced by~$P_L^{\pm,c,h}$ is very much like that in the
measure~$\BbbP_L^{\pm,J}$ conditioned on the event that the overall
magnetization~$M_L$ is near the  value~$m(h,c)L^{d}$. 
As a consequence, the asymptotic statements (e.g., the Wulff construction) 
that have been (or will be) established for the spin configurations
in the Ising model with fixed magnetization will automatically hold for the spin-marginal
of the~$P_L^{\pm,c,h}$ as well.

A particular question of interest in this paper is phase separation.
Recall that~$\mstar=\mstar(J)$ denotes the spontaneous magnetization of the  Ising
model at coupling~$J$. Then we can anticipate the following conclusions
about typical  configurations in measure~$P_L^{\pm,c,h}$:
\settowidth{\leftmargini}{(1111)}
\begin{enumerate}
\item[(1)] If $m(h,c)\ge\mstar$, then the entire system (with plus  boundary
condition) will look like the plus state of the Ising model whose external
field is  adjusted so that the overall magnetization on the scale $L^{d}$ is
roughly $m(h,c)L^{d}$.
\item[(2)] If $m(h,c)\le-\mstar$, then the system (with minus  boundary
condition) will look like the Ising minus state with similarly adjusted
external field.
\item[(3)] If $m(h,c)\in(-\mstar,\mstar)$, then, necessarily, the  system
exhibits phase separation in the sense that typical configurations feature a
large  droplet of one phase inside the other. The volume fraction taken by
the droplet is such that the  overall magnetization is near $m(h,c)L^d$. The
outer phase of the droplet agrees with the  boundary condition.
\end{enumerate}
The cases (1-2) with opposite boundary  conditions---that
is, the minus boundary conditions in~(1) and the plus boundary conditions
in~(2)---are still  as stated; the difference is that now there has to be a
large contour near the boundary  flipping to the ``correct'' boundary
condition.

\begin{remark}
\label{rem1} We have no doubt that the aforementioned conclusions  (1-3)
hold for all $d\ge2$ and all $J>\Jc$ (with a proper definition of the
\emph{droplet} in  part~(3), of course). However, the depth of
conclusion~(3) depends on the level of understanding Wulff  construction,
which is at present rather different in dimensions~$d=2$ and~$d\ge3$.
Specifically, while in $d=2$ the results of~\cite{DKS,Bob+Tim} allow us to
claim that for all $J>\Jc$  and all magnetizations
$m\in(-\mstar,\mstar)$, the system will exhibit a unique large  contour with
appropriate properties, in $d\ge3$ this statement is known to
hold~\cite{Bodineau,Cerf-Pisztora} only in ``$L^{1}$-sense'' and only for
$m\in(-\mstar,\mstar)$ which are near the endpoints. (Moreover, not all
values of~$J>\Jc$ are, in principle, permitted; cf~\cite{Bodineau2} 
for a recent improvement of these restrictions.) 
We refer to~\cite{BIV} for an overview of the situation.
\end{remark}

Notwithstanding the technical difficulties of Wulff construction, the  above
allows us to characterize the phase diagram of the model at hand.  As
indicated in  Fig.~\ref{fig1}, the
$h\le0$ and $c\ge0$ quadrant splits into three distinct parts:  The
\emph{liquid-water} region, the \emph{ice} region and the \emph{phase
separation} region, which  correspond to the situations in~(1-3),
respectively.  The boundary lines of the phase-separation  region are found
by setting
\begin{equation}
\label{separ} m(h,c)=\pm\mstar,
\end{equation}
which in light of strict monotonicity of $h\mapsto  m(h,c)$
allows us to calculate~$h$ as a function of~$c$. The solutions of
\eqref{separ}  can be obtained on the basis of the following observation:

\begin{proposition}
\label{prop2b} Let $m\in[-\mstar,\mstar]$ and $c\in[0,1]$ and define the
quantities~$q_\pm=q_\pm(m,c,\kappa)$ by~\eqref{2.8}.  Let~$h$ be
the solution to
$m(h,c)=m$. Then we have:
\begin{equation}
\label{hequals} h=\frac12\log\frac{1-q_+}{1-q_-}.
\end{equation}
In particular, there exist two continuous and  decreasing
functions
$h_\pm\colon[0,\infty)\to(-\infty,0]$ with $h_+(c)>h_-(c)$ for all
$c>0$, such that
$-\mstar<m(h,c)<\mstar$ is equivalent to $h_-(c)<h<h_+(c)$ for all $c>0$.
\end{proposition}

Proposition~\ref{prop2b} is proved at the very end of  Section~\ref{sec3.2}.
Here is an informal interpretation of this result: The quantities $q_\pm$
represent the
\emph{mole fractions} of salt in liquid-water and ice, respectively. In
mathematical terms,~$q_+$  is the probability of having a salt particle on a
given plus spin, and~$q_-$ is the corresponding  quantity for minus spins,
see~\eqref{2.7}. Formula \eqref{hequals} quantifies the shift of the
chemical potential of the solvent (which is given by $2h$ in this case) due
to the presence of  the solute. This is a manifestation of \emph{freezing
point depression}, see also Remark~\ref{rem0}. 
In the asymptotic  when $c\ll1$ we have
\begin{equation}
%\label{}
2h\approx q_--q_+.
\end{equation}
This relation, derived in standard chemistry and  physics
books under the auspicies of the ``usual approximations,'' is an essential
ingredient in the  classical analyses of colligative properties of
solutions~\cite{Moore,Landau-Lifshitz}.  Here the derivation is a direct
consequence of a microscopic (albeit simplistic) model which further  offers
the possibility of systematically calculating higher-order corrections.

\section{Proofs}
\label{sec3}\noindent 
The proofs of our main results are, more or  less,
straightforward exercises in large-deviation analysis of Gibbs
distributions. We first state  and prove a couple of technical lemmas; the
actual proofs come in Section~\ref{sec3.2}.

\subsection{Preliminaries}
\noindent
The starting point of the proof of
Theorem~\ref{thm1} (and, consequently, Theorem~\ref{thm2}) is the following
large-deviation  principle for the Ising model at zero external field:

\begin{theorem}
\label{thm3.1} 
Consider the Ising model with coupling
constant $J\in[0,\infty)$ and zero external field. Let~$\BbbP_L^{\pm,J}$ be
the corresponding (grand canonical) Gibbs measure in volume~$\Lambda_L$
and~$\pm$-boundary conditions. Then for all $m\in[-1,1]$,
\begin{equation}
\label{2.15}
\lim_{\epsilon\downarrow0}\lim_{L\to\infty}\frac1{L^d}
\log\BbbP_L^{\pm,J}\bigl(|M_L-mL^d|\le\epsilon L^d\bigr)=-\scrF_J(m),
\end{equation}
where~$M_L$ is as in \eqref{2.1} and~$\scrF_J$ is as  defined
in \eqref{2.2}.
\end{theorem}

\begin{proofsect}{Proof} The claim is considered standard, see
e.g.~\cite[Section~II.1]{Sinai}, and follows by a straightforward
application of the thermodynamic  relations between the free energy,
magnetization and external field. For completeness (and  reader's
convenience) we will provide a proof.   

Consider the function~$\phi_L(h)=\frac1{L^d}\log\E_L^{\pm,J}(e^{hM_L})$, 
where $\E_L^{\pm,J}$ is the
expectation with respect to~$\BbbP_L^{\pm,J}$, and
let~$\phi(h)=\lim_{L\to\infty}\phi_L(h)$. The limit exists by  subadditivity
arguments and is independent of the boundary condition. The
function~$h\mapsto\phi(h)$  is convex on~$\R$, real analytic (by the
Lee-Yang theorem~\cite{LY}) on~$\R\setminus\{0\}$, and hence it is
strictly convex on~$\R$. By the~$h\leftrightarrow-h$ symmetry there is a cusp at~$h=0$
whenever~$\mstar=\phi'(0^+)>0$.  It follows that for each~$m\in[\mstar,1)$
there is a unique~$\mathfrak h=\mathfrak  h(m,J)\ge0$ such that~$\phi'(\mathfrak
h)=m$, with~$\mathfrak h(m,J)$ increasing continuously from~$0$  to~$\infty$
as~$m$ increases from~$\mstar$ to~$1$. The plus-minus symmetry shows that a
similar  statement holds for the magnetizations in~$(-1,-\mstar]$.

Let~$\phi^\star$ denote the  Legendre transform of~$\phi$,
i.e.,~$\phi^\star(m)=\sup_{h\in\R}[mh-\phi(h)]$. By the above  properties
of~$h\mapsto\phi(h)$ we infer that~$\phi^\star(m)=m\mathfrak
h-\phi(\mathfrak h)$  when~$m\in(-1,-\mstar)\cup(\mstar,1)$ and~$\mathfrak
h=\mathfrak h(m,J)$, while~$\phi^\star(m)=-\phi(0)=0$
for~$m\in[-\mstar,\mstar]$. Applying the G\"artner-Ellis theorem (see
\cite[Theorem~V.6]{denHollander} 
or~\cite[Theorem~2.3.6]{Dembo-Zeitouni}), we then have \eqref{2.15}
with~$\scrF_J(m)=\phi^\star(m)$ for
all~$m\in[-1,-\mstar)\cup(\mstar,1]$---which is the set of so called exposed
points of~$\phi^\star$. Since
$\phi^\star(\pm\mstar)=0$ and the derivative of~$m\mapsto\phi^\star(m)$
is~$\mathfrak h(m,J)$, this~$\scrF_J$ is given by  the integral in
\eqref{2.2}. To prove \eqref{2.15} when~$m\in[-\mstar,\mstar]$, we must note
that  the left-hand side of
\eqref{2.15} is nonpositive and concave in~$m$. (This follows by
partitioning~$\Lambda_L$ into two parts with their own private
magnetizations and disregarding the  interaction through the boundary.)
Since~$\scrF_J(m)$ tends to zero as~$m$ tends  to~$\pm\mstar$ we thus have
that
\eqref{2.15} for~$m\in[-\mstar,\mstar]$ as well.
\end{proofsect}

\begin{remark} The ``first'' part of the G\"artner-Ellis theorem
\cite[Theorem~V.6]{denHollander} actually guarantees the following
\emph{large-deviation principle}:
\begin{equation}
%\label{}
\limsup_{L\to\infty}\frac1{L^d}
\log\BbbP_L^{\pm,J}\Bigl(\frac{M_L}{L^d}\in\CC\Bigr)\le-\inf_{m\in\CC}\phi^\star(m)
\end{equation}
for any closed set~$\CC\subset\R$ while
\begin{equation}
%\label{}
\liminf_{L\to\infty}\frac1{L^d}
\log\BbbP_L^{\pm,J}\Bigl(\frac{M_L}{L^d}\in\OO\Bigr)\ge-\inf_{m\in\OO\smallsetminus[-\mstar,\mstar]}\phi^\star(m)
\end{equation}
for any open set~$\OO\subset\R$.
(Here~$\phi^\star(m)=\scrF_J(m)$ for~$m\in[-1,1]$ and $\phi^\star(m)=\infty$
otherwise.) The above proof follows by  specializing to
$\epsilon$-neighborhoods of a given~$m$ and letting~$\epsilon\downarrow0$.
The~$m\in[-\mstar,\mstar]$ cases---i.e, the non-exposed points---have  to be
dealt with separately.
\end{remark}

The above is the core of our proof of Theorem~\ref{thm1}. The next  step
will be to bring the quantities~$c$ and~$h$ into play. This, as we shall
see, is easily  done if we condition on the total magnetization. (The cost
of this conditioning will be estimated  by \eqref{2.15}.) Indeed, as a
result of the absence of salt-salt interaction, the conditional  measure can
be rather precisely characterized. Let us recall the definition of the
quantity~$N_L$ from \eqref{1.9} which represents the total amount of salt in
the system. For any spin
configuration~$\sigma=(\sigma_x)\in\{-1,1\}^{\Lambda_L}$ and any salt
configuration~$\sS=(\sS_x)\in\{0,1\}^{\Lambda_L}$, let us introduce  the
quantity
\begin{equation}
%\label{}
Q_L=Q_L(\sigma,\sS)=\sum_{x\in\Lambda_L}\sS_x\frac{1+\sigma_x}2
\end{equation}
representing the total amount of salt ``on the plus  spins.''
Then we have:

\begin{lemma}
\label{lemma3.2} For any fixed spin
configuration~$\bar\sigma=(\bar\sigma_x)\in\{-1,1\}^{\Lambda_L}$, all  salt
configurations
$(\sS_x)\in\{0,1\}^{\Lambda_L}$ with the same~$N_L$ and~$Q_L$ have  the same
probability in the conditional
measure~$P_L^{\pm,c,h}(\ccdot|\sigma=\bar\sigma)$. Moreover, for
any~$\barsS=(\barsS_x)\in\{0,1\}^{\Lambda_L}$ with $N_L=\lfloor
cL^d\rfloor$ and for any~$m\in[-1,1]$,
\begin{multline}
\label{3.3}  
\quad 
P_L^{\pm,c,h}\bigl(\,\barsS\,\text{\rm~occurs},\,M_L=\lfloor
mL^d\rfloor\bigr)
\\=\frac1{Z_L} \E_L^{\pm,J}\bigl(e^{\kappa Q_L(\sigma,\barsS)+hM_L(\sigma)}
\1_{\{M_L(\sigma)=\lfloor mL^d\rfloor\}}\bigr),
\quad
\end{multline}
where the normalization constant is given by
\begin{equation}
\label{3.4}  Z_L=\sum_{\sS'\in\{0,1\}^{\Lambda_L}}\1_{\{N_L(\sS')=\lfloor
cL^d\rfloor\}}\,
\E_L^{\pm,J}\bigl(e^{\kappa Q_L(\sigma,\sS')+hM_L(\sigma)}\bigr).
\end{equation}
Here~$\E_L^{\pm,J}$ is the expectation with respect
to~$\BbbP_L^{\pm,J}$.
\end{lemma}

\begin{proofsect}{Proof} The fact that all salt configurations with
given~$N_L$ and~$Q_L$ have the same probability
in~$P_L^{\pm,c,h}(\ccdot|\sigma=\bar\sigma)$ is  a consequence of the
observation that the salt-dependent part of the Hamiltonian
\eqref{2.1a} depends only on~$Q_L$. The relations \twoeqref{3.3}{3.4} follow
by a straightforward rewrite  of the overall Boltzmann weight.
\end{proofsect}

The characterization of the conditional measure
$P_L^{\pm,c,h}(\ccdot|M_L=\lfloor mL^d\rfloor)$ from Lemma~\ref{lemma3.2}
allows us to explicitly evaluate the  configurational entropy carried by the
salt. Specifically, given a spin
configuration~$\sigma=(\sigma_x)\in\{-1,1\}^{\Lambda_L}$ and
numbers~$\theta,c\in(0,1)$, let
\begin{equation}
%\label{}
\AA_L^{\theta,c}(\sigma)=\bigl\{(\sS_x)\in\{0,1\}^{\Lambda_L}\colon
N_L=\lfloor cL^d\rfloor,\,Q_L=\lfloor\theta cL^d\rfloor\bigr\}.
\end{equation}
The salt entropy is then the rate of exponential  growth of
the size of~$\AA_L^{\theta,c}(\sigma)$ which can be related to the quantity
$\Xi(m,\theta;c)$ from
\eqref{2.3} as follows:

\begin{lemma}
\label{lemma3.3} For each~$\epsilon'>0$ and each~$\eta>0$ there exists a  number~$L_0<\infty$
such that the following is true for any $\theta,c\in(0,1)$, any~$m\in(-1,1)$
that obey $|m|\le1-\eta$,
\begin{equation}
\label{3.5}
\frac{2\theta c}{1+m}\le1-\eta\quad\text{and}\quad\frac{2(1-\theta)c}{1-m}\le1-\eta,
\end{equation}
and any~$L\ge L_0$: If
$\sigma=(\sigma_x)\in\{-1,1\}^{\Lambda_L}$ is a spin configuration with
$M_L(\sigma)=\lfloor mL^d\rfloor$, then
\begin{equation}
\label{3.6}
\biggl|\frac{\log|\AA_L^{\theta,c}(\sigma)|}{L^d}-\Xi(m,\theta;c)\biggr|\le\epsilon'.
\end{equation}
\end{lemma}

\begin{proofsect}{Proof} 
We want to distribute~$N_L=\lfloor  cL^d\rfloor$
salt particles over~$L^d$ positions, such that exactly $Q_L=\lfloor\theta
cL^d\rfloor$ of them  land on~$\frac12(L^d+M_L)$ plus sites and~$N_L-Q_L$ on
$\frac12(L^d-M_L)$ minus sites. This can be done in
\begin{equation}
\label{3.8a}
|\AA_L^{\theta,c}(\sigma)|=\binom{\frac12(L^d+M_L)}{Q_L}\binom{\frac12(L^d-M_L)}{N_L-Q_L}
\end{equation}
number of ways. Now all quantities scale proportionally to~$L^d$
which, applying Stirling's formula, shows that the first term is within,
say,~$e^{\pm  L^d\epsilon'/2}$ multiples of
\begin{equation}
%\label{}
\exp\biggl\{-L^d\frac{1+m}2\scrS\Bigl(\frac{2\theta c}{1+m}\Bigr)\biggr\}
\end{equation}
once~$L\ge L_0$, with~$L_0$ depending only  on~$\epsilon'$.
A similar argument holds also for the second term with~$\theta$
replaced by~$1-\theta$  and~$m$ by~$-m$. Combining these expressions we get
that~$|\AA_L^{\theta,c}(\sigma)|$ is  within~$e^{\pm L^d\epsilon'}$
multiples of~$\exp\{L^d\Xi(m,\theta;c)\}$ once~$L$ is sufficiently large.
\end{proofsect}

For the proof of Theorem~\ref{thm2}, we will also need an estimate on  how
many salt configurations in~$\AA_L^{\theta,c}(\sigma)$ take given values in
a finite subset~$\Lambda\subset\Lambda_L$. To that extent, for
each~$\sigma\in\{-1,1\}^{\Lambda_L}$ and
each~$\barsS_\Lambda\in\{0,1\}^\Lambda$ we will define the quantity
\begin{equation}
\label{3.7}
R_{\Lambda,L}^{\theta,c}(\sigma,\barsS_\Lambda)=
\frac{|\{\barsS\in\AA_L^{\theta,c}(\sigma)\colon
\sS_\Lambda=\barsS_\Lambda\}|}{|\AA_L^{\theta,c}(\sigma)|}.
\end{equation}
As a moment's thought
reveals,~$R_{\Lambda,L}^{\theta,c}(\sigma,\barsS_\Lambda)$ can be
interpreted as the probability  that~$\{\sS_\Lambda=\barsS_\Lambda\}$ occurs
in (essentially) any homogeneous product measure
on~$\sS=(\sS_x)\in\{0,1\}^{\Lambda_L}$ conditioned to have~$N_L(\sS)=\lfloor
cL^d\rfloor$  and~$Q_L(\sigma,\sS)=\lfloor\theta cL^d\rfloor$. It is
therefore not surprising that, for spin configurations~$\sigma$ with given
magnetization,~$R_{\Lambda,L}^{\theta,c}(\sigma,\cdot)$ will tend to  a
product measure on~$\sS_\Lambda\in\{0,1\}^\Lambda$. A precise
characterization of  this limit is as follows:
\begin{lemma}
\label{lemma3.4} For each~$\epsilon>0$, each~$K\ge1$ and  each~$\eta>0$
there exists~$L_0<\infty$ such that the following holds for all~$L\ge L_0$,
all~$\Lambda\subset\Lambda_L$ with~$|\Lambda|\le K$, all~$m$
with~$|m|\le1-\eta$ and  all~$\theta,c\in[\eta,1-\eta]$ for which
\begin{equation}
\label{3.5a} 
p_+=\frac{2\theta c}{1+m}\quad\text{and}\quad
p_-=\frac{2(1-\theta)c}{1-m}
\end{equation}
satisfy~$p_\pm\in[\eta,1-\eta]$:
If~$\sigma=(\sigma_x)\in\{-1,1\}^{\Lambda_L}$ is a spin configuration such
that~$M_L(\sigma)=\lfloor mL^d\rfloor$
and~$\barsS_\Lambda\in\{0,1\}^\Lambda$ is a salt configuration
in~$\Lambda$, then
\begin{equation}
%\label{}
\Bigl|R_{\Lambda,L}^{\theta,c}(\sigma,\sS_\Lambda)-
\prod_{x\in
\Lambda}\bigl\{p_{\sigma_x}\delta_1(\barsS_x)+(1-p_{\sigma_x})\delta_0(\barsS_x)\bigr\}\Bigr|\le\epsilon.
\end{equation}
\end{lemma}

\begin{proofsect}{Proof} We will expand on the argument from
Lemma~\ref{lemma3.3}. Indeed, from~\eqref{3.8a} we have an expression for
the denominator in
\eqref{3.7}. As to the numerator, introducing the quantities
\begin{equation}
%\label{}
M_\Lambda=\sum_{x\in\Lambda}\sigma_x,
\quad N_\Lambda=\sum_{x\in\Lambda}\sS_x,
\quad Q_\Lambda=\sum_{x\in\Lambda}\sS_x\frac{1+\sigma_x}2,
\end{equation}
and the shorthand
\begin{equation}
%\label{}
D=D_{r,r',s,s'}(\ell,\ell',q,q')=\frac{\displaystyle\binom{r-\ell}{s-q}\binom{r'-\ell'}{s'-q'}}
{\displaystyle\binom rs\binom{r'}{s'}},
\end{equation}
the same reasoning as we used to prove \eqref{3.8a} allows us
to write the object~$R_{\Lambda,L}^{\theta,c}(\sigma,\sS_\Lambda)$
as $D_{r,r',s,s'}(\ell,\ell',q,q')$, where the various parameters are as
follows: The quantities
\begin{equation}
%\label{}
r=\frac{L^d+M_L}2\quad\text{and}\quad r'=\frac{L^d-M_L}2
\end{equation}
represent the total number of pluses and minuses in  the
system, respectively,
\begin{equation}
%\label{}
s=Q_L\quad\text{and}\quad s'=N_L-Q_L
\end{equation}
are the numbers of salt particles on pluses and  minuses,
and, finally,
\begin{equation}
%\label{}
\ell=\frac{|\Lambda|+M_\Lambda}2,\quad\ell'=\frac{|\Lambda|-M_\Lambda}2,\quad
q=Q_\Lambda\quad\text{and}\quad q'=N_\Lambda-Q_\Lambda
\end{equation}
are the corresponding quantities for the  volume~$\Lambda$,
respectively.

Since~\eqref{3.5a} and the restrictions on~$|m|\le1-\eta$ and
$\theta,c\in[\eta,1-\eta]$ imply that~$r$,~$r'$,~$s$,~$s'$,~$r-s$
and~$r'-s'$ all scale proportionally  to~$L^d$, uniformly in~$\sigma$
and~$\sS_\Lambda$, while~$\ell$ and~$\ell'$ are bounded
by~$|\Lambda|$---which by our assumption is less than~$K$---we are in
a regime where it makes sense to seek an
asymptotic form of quantity~$D$. Using the bounds
\begin{equation}
%\label{}
a^be^{-b^2/a}\le\frac{(a+b)!}{a!}\le a^be^{b^2/a},
\end{equation}
which are valid for all integers~$a$ and~$b$ with~$|b|\le a$, we
easily find that
\begin{equation}
%\label{}
D=\Bigl(\frac sr\Bigr)^\ell
\Bigl(1-\frac sr\Bigr)^{\ell-q}
\Bigl(\frac{s'}{r'}\Bigr)^{\ell'}
\Bigl(1-\frac{s'}{r'}\Bigr)^{\ell'-q'}+o(1),
\qquad L\to\infty.
\end{equation}
Since~$s/r\to p_+$ and~$s'/r'\to p_-$ as~$L\to\infty$, while~$\ell$, $q$, $\ell'$ and~$q'$
stay bounded, the desired claim follows by taking~$L$ sufficiently large.
\end{proofsect}

The reader may have noticed that, in most of our previous arguments,
$\theta$ and~$m$ were restricted to be away from the boundary values. To
control the  situation near the boundary values, we have to prove the
following claim:

\begin{lemma}
\label{lemma3.5} For each~$\epsilon\in(0,1)$ and each~$L\ge1$,
let~$\EE_{L,\epsilon}$ be the event
\begin{multline}
\label{3.22}
\quad
\EE_{L,\epsilon}=\bigl\{|M_L|\le(1-\epsilon)L^d\bigr\}
\\
\cap\bigl\{\epsilon
\tfrac12(L^d+M_L)\le Q_L\le(1-\epsilon)\tfrac12(L^d+M_L)\bigr\}.
\quad
\end{multline}
Then for each~$c\in(0,1)$ and each~$h\in\R$ there exists
an~$\epsilon>0$ such that
\begin{equation}
%\label{}
\limsup_{L\to\infty}\frac1{L^d}\log
P_L^{\pm,c,h}\bigl(\EE_{L,\epsilon}^\cc)<0.
\end{equation}
\end{lemma}

\begin{proofsect}{Proof} We will split the complement of
$\EE_{L,\epsilon}$ into four events and prove the corresponding estimate for
each of them. We begin with the event~$\{M_L\le-(1-\epsilon)L^d\}$. The main
tool will be stochastic  domination by a product measure. Consider the usual
partial order on spin configurations defined by putting~$\sigma\prec\sigma'$
whenever~$\sigma_x\le\sigma'_x$ for all~$x$. Let
\begin{equation}
\label{3.24a}
\lambda=\inf_{L\ge1}\min_{x\in\Lambda_L}\,\min_{\begin{subarray}{c}
\sigma'\in\{-1,1\}^{\Lambda_L\smallsetminus\{x\}}\\\barsS\in\{-1,1\}^{\Lambda_L}
\end{subarray}} P_L^{\pm,c,h}(\sigma_x=1|\sigma',\barsS)
\end{equation}
be the conditional probability that~$\sigma_x=+1$ occurs
given a spin configuration~$\sigma'$ in $\Lambda_L\setminus\{x\}$ and a salt configuration~$\barsS$ in~$\Lambda_L$, optimized over all~$\sigma'$, $\barsS$ 
and also~$x\in\Lambda_L$ and the system size. 
Since~$P_L^{\pm,c,h}(\sigma_x=1|\sigma',\barsS)$  reduces to (the
exponential of) the local interaction between~$\sigma_x$ and its ultimate
neighborhood, we have~$\lambda>0$.   

Using standard arguments it now follows
that the  spin marginal of~$P_L^{\pm,c,h}$ stochastically dominates the
product  measure~$\BbbP_\lambda$ defined
by~$\BbbP_\lambda(\sigma_x=1)=\lambda$ for all~$x$. In particular, we have
\begin{equation}
%\label{}
P_L^{\pm,c,h}\bigl(M_L\le
-(1-\epsilon)L^d\bigr)\le\BbbP_\lambda\bigl(M_L\le -(1-\epsilon)L^d\bigr).
\end{equation}
Let $\epsilon<2\lambda$.
Then~$\lambda-(1-\lambda)$---namely, the expectation of~$\sigma_x$ with
respect to~$\BbbP_\lambda$---exceeds the negative  of~$(1-\epsilon)$ and so
Cram\'er's theorem (see \cite[Theorem~I.4]{denHollander} or
\cite[Theorem~2.1.24]{Dembo-Zeitouni}) implies that the probability  on the
right-hand side decays to zero exponentially in~$L^d$, i.e.,
\begin{equation}
%\label{}
\limsup_{L\to\infty}\frac1{L^d}\log\BbbP_\lambda\bigl(M_L\le
-(1-\epsilon)L^d\bigr)<0.
\end{equation}
The opposite side of the interval of magnetizations, namely, the event $\{M_L\ge(1-\epsilon)L^d\}$, is handled  analogously
(with~$\lambda$ now focusing on~$\sigma_x=-1$ instead of~$\sigma_x=1$).

The remaining two events, marking when~$Q_L$ is either less  than~$\epsilon$
or larger than~$(1-\epsilon)$ times the total number of plus spins, are
handled  using a similar argument combined with standard convexity estimates. 
Let us consider the event~$\{Q_L\le\epsilon  L^d\}$---which
contains the event~$\{Q_L\le\epsilon\frac12(M_L+L^d)\}$---and let us emphasize  the
dependence of the underlying probability distribution on~$\kappa$ by writing~$P_L^{\pm,c,h}$ as~$\BbbP_\kappa$.
Let~$\E_\kappa$ denote the expectation with respect to~$\BbbP_\kappa$ and note
that~$\E_\kappa(f)=\E_0(fe^{\kappa  Q_L})/\E_0(e^{\kappa Q_L})$. We begin by
using the Chernoff bound to get
\begin{equation}
%\label{}
\BbbP_\kappa(Q_L\le\epsilon L^d)\le e^{a\epsilon
L^d}\E_\kappa(e^{-aQ_L})=\frac{e^{a\epsilon
L^d}}{\E_{\kappa-a}(e^{aQ_L})},\qquad a\ge0.
\end{equation}
A routine application of Jensen's inequality gives us
\begin{equation}
%\label{}
\BbbP_\kappa(Q_L\le\epsilon L^d)\le \exp\Bigl\{a\bigl(\epsilon
L^d-\E_{\kappa-a}(Q_L)\bigl)\Bigr\}.
\end{equation}
It thus suffices to prove that there  exists a~$\kappa'<\kappa$
such that
$\inf_{L\ge1}\frac1{L^d}\E_{\kappa'}(Q_L)$ is positive. (Indeed, we take~$\epsilon$ to be strictly less than this number and
set~$a=\kappa-\kappa'$ to observe that the right-hand side decays
exponentially in~$L^d$.) To show this we  write $\E_{\kappa'}(Q_L)$ as the sum
of~$\BbbP_{\kappa'}(\sigma_x=1,\sS_x=1)$ over  all~$x\in\Lambda_L$. Looking
back at
\eqref{3.24a}, we then have
$\BbbP_{\kappa'}(\sigma_x=1,\sS_x=1)\ge\lambda\BbbP_{\kappa'}(\sS_x=1)$,
where~$\lambda$ is now evaluated for~$\kappa'$, and so
\begin{equation}
%\label{}
\E_{\kappa'}(Q_L)\ge\lambda\sum_{x\in\Lambda_L}\BbbP_{\kappa'}(S_x=1)
=\lambda\E_{\kappa'}(N_L)\approx\lambda cL^d.
\end{equation}
Thus, once~$\lambda c>\epsilon$, the
probability~$\BbbP_\kappa(Q_L\le\epsilon L^d)$ decays exponentially
in~$L^d$.

As to the complementary event,
$\{Q_L\ge(1-\epsilon)\frac12(M_L+L^d)\}$, we note that this is contained
in~$\{H_L\le\epsilon L^d\}$, where~$H_L$ counts the number  of plus spins
with no salt on it. Since we still have~$\E_\kappa(f)=\E_0(fe^{-\kappa
H_L})/\E_0(e^{-\kappa H_L})$, the proof boils down to the same argument as
before.
\end{proofsect}

\subsection{Proofs of Theorems~\ref{thm1} and~\ref{thm2}}
\label{sec3.2}\noindent On the basis of the above observations, the  proofs
of our main theorems are easily concluded. However, instead of
Theorem~\ref{thm1} we will  prove a slightly stronger result of which the
large-deviation part of Theorem~\ref{thm1} is an  easy corollary.

\begin{theorem}
\label{thm3.5} 
Let~$J>0$ and~$\kappa\ge0$ be fixed. For
each~$c,\theta\in(0,1)$, each~$h\in\R$ and each~$m\in(-1,1)$,
let~$\BB_{L,\epsilon}=\BB_{L,\epsilon}(m,c,\theta)$ be the set of all
$(\sigma,\sS)\in\{-1,1\}^{\Lambda_L}\times\{0,1\}^{\Lambda_L}$ for  which
$|M_L-mL^d|\le\epsilon L^d$ and $|Q_L-\theta cL^d|\le\epsilon L^d$ hold.
Then
\begin{equation}
\label{3.10}
\lim_{\epsilon\downarrow0}\lim_{L\to\infty}\frac{\log
P_L^{\pm,c,h}(\BB_{L,\epsilon})}{L^d}=
-\scrG_{h,c}(m,\theta)+\inf_{\begin{subarray}{c}
m'\in(-1,1)\\\theta'\in[0,1]
\end{subarray}}
\scrG_{h,c}(m',\theta'),
\end{equation}
where~$\scrG_{h,c}(m,\theta)$ is as in \eqref{Phi2}.
\end{theorem}

\begin{proofsect}{Proof} Since the size of the
set~$\AA_L^{\theta,c}(\sigma)$ depends only on the overall magnetization,  let~$A_L^{\theta,c}(m)$ denote this size for the
configurations~$\sigma$ with~$M_L(\sigma)=\lfloor
mL^d\rfloor$. First we note that, by Lemma~\ref{lemma3.2},
\begin{equation}
\label{3.7a} 
P_L^{\pm,c,h}\bigl(Q_L=\lfloor\theta cL^d\rfloor,\,M_L=\lfloor
mL^d\rfloor\bigr)=\frac{K_L(m,\theta)}{Z_L}
\end{equation}
where
\begin{equation}
\label{3.8} 
K_L(m,\theta)=A_L^{\theta,c}(m)\,e^{h\lfloor
mL^d\rfloor+\kappa\lfloor\theta cL^d\rfloor}\,
\BbbP_L^{\pm,J}\bigl(M_L=\lfloor mL^d\rfloor\bigr).
\end{equation}
Here~$Z_L$ is the normalization constant from
\eqref{3.4} which in the present formulation can also be interpreted as the
sum of $K_L(m,\theta)$  over the relevant (discrete) values of~$m$
and~$\theta$.

Let~$K_{L,\epsilon}(m,\theta)$ denote the sum of~$K_L(m',\theta)$  over
all~$m'$ and~$\theta'$ for which~$m'L^d$ and~$\theta'cL^d$ are integers and
$|m'-m|\le\epsilon$  and~$|\theta'c-\theta c|\le\epsilon$. (This is exactly
the set of magnetizations and  spin-salt overlaps contributing to the
set~$\BB_{L,\epsilon}$.) Applying \eqref{2.15} to extract the  exponential
behavior of the last probability in \eqref{3.8}, and using \eqref{3.6} to do
the same  for the quantity
$A_L^{\theta,c}(m)$, we get
\begin{equation}
\label{3.9}
\Bigl|\frac{\log
K_{L,\epsilon}(m,\theta)}{L^d}+\scrG_{h,c}(m,\theta)\Bigr|\le\epsilon+\epsilon',
\end{equation}
where~$\epsilon'$ is as in \eqref{3.6}. As a  consequence of
the above estimate we have
\begin{equation}
%\label{}
\lim_{\epsilon\downarrow0}\lim_{L\to\infty}\frac{\log
K_{L,\epsilon}(m,\theta)}{L^d}=-\scrG_{h,c}(m,\theta)
\end{equation}
for any~$m\in(-1,1)$ and any~$\theta\in(0,1)$.

Next we will attend to the denominator in \eqref{3.7a}.  Pick~$\delta>0$ and
consider the set
\begin{equation}
%\label{}
\MM_\delta=\bigl\{(m,\theta)\colon
|m|\le1-\delta,\,\delta\le\theta\le1-\delta\bigr\}.
\end{equation}
We will write~$Z_L$ as a sum of two terms,
$Z_L=Z_L^{(1)}+Z_L^{(2)}$, with~$Z_L^{(1)}$ obtained by summing
$K(m,\theta)$ over the  admissible~$(m,\theta)\in\MM_\delta$ and~$Z_L^{(2)}$
collecting the remaining terms. By Lemma~\ref{lemma3.5} we know
that~$Z_L^{(2)}/Z_L$ decays exponentially in~$L^d$ and so the  decisive
contribution to~$Z_L$ comes from~$Z_L^{(1)}$. Assuming
that~$\epsilon\ll\delta$, let us  cover $\MM_\delta$ by finite number of
sets of the
form~$[m_\ell'-\epsilon,m_\ell'+\epsilon]\times[\theta_\ell'-\epsilon,\theta_\ell'+\epsilon]$,
where~$m_\ell'$ and~$\theta_\ell'$ are such that~$m_\ell'L^d$
and~$\theta_\ell'cL^d$ are integers. Then~$Z_L^{(1)}$ can be bounded as in
\begin{equation}
%\label{}
\max_\ell K_{L,\epsilon}(m_\ell',\theta_\ell')\le Z_L^{(1)}\le\sum_\ell
K_{L,\epsilon}(m_\ell',\theta_\ell').
\end{equation}
Moreover, the right-hand side is bounded by the left-hand side times
a polynomial in~$L$.
Taking logarithms, dividing by~$L^d$, taking the
limit~$L\to\infty$, refining the cover and applying the continuity
of~$(m,\theta)\mapsto\scrG_{h,c}(m,\theta)$ allows us to conclude that
\begin{equation}
\label{3.15}
\lim_{L\to\infty}\frac{\log Z_L}{L^d}=
-\inf_{m\in(-1,1)}\inf_{\theta\in[0,1]}\scrG_{h,c}(m,\theta).
\end{equation}
Combining these observations, \eqref{2.4} is proved.
\end{proofsect}

\begin{proofsect}{Proof of Theorem~\ref{thm1}} The conclusion
\eqref{2.4} follows from
\eqref{3.10} by similar arguments that prove \eqref{3.15}. The only
remaining thing to prove is the strict convexity of~$m\mapsto G_{h,c}(m)$ and
continuity and  monotonicity of its minimizer. First we note
that~$\theta\mapsto\scrG_{h,c}(m,\theta)$ is strictly convex  on the set
of~$\theta$ where it is finite, which is a simple consequence of the strict
convexity  of~$p\mapsto \scrS(p)$. Hence, for each~$m$, there is a
unique~$\theta=\theta(m)$ which
minimizes~$\theta\mapsto\scrG_{h,c}(m,\theta)$.   

Our next goal is to  show
that, for~$\kappa c>0$, the solution~$\theta=\theta(m)$ will satisfy the inequality
\begin{equation}
\label{3.19}
\theta>\frac{1+m}2.
\end{equation}
(A heuristic reason for this is  that~$\theta=\frac{1+m}2$
corresponds to the situation when the salt is distributed independently of
the  underlying spins. This is the dominating strategy for~$\kappa=0$;
once~$\kappa>0$ it is clear that  the fraction of salt on plus spins
\emph{must} increase.) 
A formal proof runs as follows: We first note that~$m\mapsto\theta(m)$ 
solves for~$\theta$ from the equation
\begin{equation}
\label{3.20}
\frac\partial{\partial\theta}\Xi(m,\theta;c)=-\kappa c,
\end{equation}
where~$\Xi(m,\theta;c)$ is as in \eqref{2.3}.
But~$\theta\mapsto\Xi(m,\theta;c)$ is strictly concave and its derivative
vanishes  at~$\theta=\frac12(1+m)$. Therefore, for~$\kappa c>0$ the
solution~$\theta=\theta(m)$ of \eqref{3.20} must obey
\eqref{3.19}.  

Let~$\VV$ be the set of~$(m,\theta)\in(-1,1)\times(0,1)$
for which \eqref{3.19} holds  and note that~$\VV$ is convex. A standard
second-derivative calculation now shows  that~$\scrG_{h,c}(m,\theta)$ is
strictly convex on~$\VV$. (Here we actually differentiate the function
$\scrG_{h,c}(m,\theta)-\scrF_J(m)$---which is twice differentiable on  the set
where it is finite---and then use the known convexity of~$\scrF_J(m)$. The
strict  convexity is violated on the line~$\theta=\frac12(1+m)$
where~$(m,\theta)\mapsto\scrG_{h,c}(m,\theta)$ has a flat piece
for~$m\in[-\mstar,\mstar]$.) Now, since~$\theta(m)$
minimizes~$\scrG_{h,c}(m,\theta)$ for a given~$m$, the strict convexity
of~$\scrG_{h,c}(m,\theta)$ on~$\VV$  implies that for
any~$\lambda\in(0,1)$,
\begin{multline}
G_{h,c}\bigl(\lambda m_1+(1-\lambda)m_2\bigr)
\\ 
\begin{aligned}
&\le
\scrG_{h,c}\bigl(\lambda
m_1+(1-\lambda)m_2,\lambda\theta(m_1)+(1-\lambda)\theta(m_2)\bigr)
\\ &<\lambda\scrG_{h,c}\bigl(m_1,\theta(m_1)\bigr)+(1-\lambda)
\scrG_{h,c}\bigl(m_2,\theta(m_2)\bigr)
\\ &=\lambda G_{h,c}(m_1)+(1-\lambda)G_{h,c}(m_2).
\end{aligned}
\end{multline}
Hence,~$m\mapsto G_{h,c}(m)$ is also strictly convex.  The
fact that~$G'(m)$ diverges as~$m\to\pm1$ is a consequence of the
corresponding property of the function~$m\mapsto\scrF_J(m)$ and the fact
that the rest  of~$\scrG_{h,c}$ is convex in~$m$.

As a consequence of
strict convexity and the abovementioned ``steepness''  at the boundary of
the interval~$(-1,1)$, the function~$m\mapsto G_{h,c}(m)$ has a unique
minimizer for each~$h\in\R$ and~$c>0$, as long as the quantities from \eqref{3.5a}
satisfy~$p_\pm<1$. The minimizer is  automatically continuous in~$h$ and is manifestly
non-decreasing. Furthermore, the continuity  of~$G_{h,c}$ in~$c$ allows us
to conclude that~$\theta(m)$ is also continuous in~$c$. What is  left of the
claims is the \emph{strict} monotonicity of~$m$ as a function
of~$h$.  Writing~$G_{h,c}(m)$ as~$-hm+g(m)$ and noting that~$g$ is
continuously differentiable on~$(-1,1)$, the  minimizing~$m$ satisfies
\begin{equation}
%\label{}
g'(m)=h.
\end{equation}
But~$g(m)$ is also strictly convex and so~$g'(m)$ is  strictly
increasing. It follows that~$m$ has to be strictly increasing with~$h$.
\end{proofsect}

Theorem~\ref{thm3.1} has the following simple consequence that is  worth
highlighting:

\begin{corollary}
\label{cor3.6} For given~$h\in\R$ and~$c\in(0,1)$, let~$(m,\theta)$  be the
minimizer of~$\scrG_{h,c}(m,\theta)$. Then for all~$\epsilon>0$,
\begin{equation}
%\label{}\
\lim_{L\to\infty} P_L^{\pm,c,h}\bigl(|Q_L-\theta cL^d|\ge\epsilon
L^d\text{\rm\ or }|M_L-mL^d|\ge\epsilon L^d\bigr)=0.
\end{equation}
\end{corollary}

\begin{proofsect}{Proof} On the basis of \eqref{3.10} and the fact
that~$\scrG_{h,c}(m,\theta)$ has a unique minimizer, a covering
argument---same as used to prove
\eqref{3.15}---implies that the probability on the left-hand side decays to
zero exponentially in~$L^d$.
\end{proofsect}

Before we proceed to the proof of our second main theorem, let us  make an
observation concerning the values of~$p_\pm$ at the minimizing~$m$
and~$\theta$:
\begin{lemma}
\label{lemma3.7} Let~$h\in\R$ and~$c\in(0,1)$ be fixed and  let~$(m,\theta)$
be the minimizer of~$\scrG_{h,c}(m,\theta)$. Define the
quantities~$q_\pm=q_\pm(m,c,\kappa)$ by \eqref{2.8}
and~$p_\pm=p_\pm(m,\theta,c)$ by \eqref{3.5a}. Then
\begin{equation}
\label{3.26} 
q_+=p_+\quad\text{and}\quad q_-=p_-.
\end{equation}
Moreover,~$q_\pm$ are then related to~$h$ via \eqref{hequals}
whenever~$m\in[-\mstar,\mstar]$.
\end{lemma}

\begin{proofsect}{Proof} First let us ascertain that~$q_\pm$ are well
defined from equations
\eqref{2.8}. We begin by noting that the set of possible values
of~$(q_+,q_-)$ is the unit square~$[0,1]^2$. As is easily shown, the first
equation in
\eqref{2.8} corresponds to an increasing curve in~$[0,1]^2$ connecting the
corners~$(0,0)$  and~$(1,1)$. On the other hand, the second equation in
\eqref{2.8} is a straight line with negative slope  which by the fact
that~$c<1$ intersects both the top and the right side of the square.  It
follows that these curves intersect at a single point---the unique solution
of \eqref{2.8}. 

Next we will derive equations that~$p_\pm$ have to satisfy.
Let~$(m,\theta)$ be the unique minimizer of~$\scrG_{h,c}(m,\theta)$. 
The partial derivative with respect  to~$\theta$ yields
\begin{equation}
%\label{}
c\bigl(\scrS'(p_+)-\scrS'(p_-)\bigr)=\kappa c
\end{equation}
and from the very definition of~$p_\pm$ we
have
\begin{equation}
%\label{}
\frac{1+m}2p_++\frac{1-m}2p_-=c.
\end{equation}
Noting that~$\scrS'(p)=\log\frac p{1-p}$, we now see
that~$p_\pm$ satisfies the same equations as~$q_\pm$ and so, by the above
uniqueness argument,
\eqref{3.26} must hold.

To prove relation \eqref{hequals}, let us also consider the  derivative
of~$\scrG_{h,c}(m,\theta)$ with respect to~$m$. For solutions
in~$[-\mstar,\mstar]$ we can  disregard the~$\scrF_J$ part of the function
(because its vanishes along with its derivative  throughout this interval),
so we have
\begin{equation}
%\label{}
h=-\frac\partial{\partial m}\Xi(m,\theta;c).
\end{equation}
A straightforward calculation then yields \eqref{hequals}.
\end{proofsect}

Now we are ready to prove our second main result:

\begin{proofsect}{Proof of Theorem~\ref{thm2}} 
The crucial technical  step
for the present proof has already been established in Lemma~\ref{lemma3.2}.
In order to  plug into the latter result, let us note that the sum
of~$e^{\kappa Q_L(\sigma,\sS)}$ over all salt
configurations~$\sS=(\sS_x)\in\{0,1\}^{\Lambda_L}$ with~$N_L=\lfloor
cL^d\rfloor$ is a number depending only on the total
magnetization~$M_L=M_L(\sigma)$.  Lemma~\ref{lemma3.2} then implies
\begin{multline}
%\label{}
\quad
P_L^{\pm,c,h}\bigl(\AA\times\{0,1\}^{\Lambda_L}\cap\{M_L=\lfloor mL^d\rfloor\}\bigr) 
\\
=\omega_L(m)\,\BbbP_L^{\pm,J}\bigl(\AA\cap\{M_L=\lfloor
mL^d\rfloor\}\bigr)
\quad\end{multline}
where~$\omega_L(m)$ is a positive number depending on~$m$,
the parameters~$c$,~$h$,~$J$ and the boundary condition~$\pm$ but not on
the event~$\AA$. Noting that~$\rho_L^\pm$ is simply the distribution of the
random  variables~$M_L/L^d$ in measure~$P_L^{\pm,c,h}$, this proves~\eqref{2.6}.  

In order to prove  the assertion \eqref{2.7}, we
let~$\bar\sigma\in\{0,1\}^{\Lambda_L}$, pick~$\Lambda\subset\Lambda_L$ and
fix~$\barsS\in\{0,1\}^\Lambda$. Since Lemma~\ref{lemma3.2} guarantees that,
given~$\{\sigma=\bar\sigma\}$, all salt configurations with  fixed~$Q_L$ and
concentration~$c$ have the same probability
in~$P_L^{\pm,c,h}(\ccdot|\sigma=\bar\sigma)$, we have
\begin{equation}
\label{3.33}
P_L^{\pm,c,h}\bigl(\sS_\Lambda=\barsS_\Lambda,\,\sS\in\AA_L^{\theta,c}(\bar\sigma)\big|\sigma=\bar\sigma\bigr)
=R_{\Lambda,L}^{\theta,c}(\bar\sigma,\sS_\Lambda),
\end{equation}
where~$R_{\Lambda,L}^{\theta,c}$ is defined in
\eqref{3.7}. Pick~$\eta>0$ and assume, as in Lemma~\ref{lemma3.4},
that~$c\in[\eta,1-\eta]$,
$\theta\in[\eta,1-\eta]$ and~$M_L(\bar\sigma)=\lfloor mL^d\rfloor$ for
some~$m$  with~$|m|\le1-\eta$. Then the aforementioned lemma tells us that
$R_{\Lambda,L}^{\theta,c}(\bar\sigma,\cdot)$ is within~$\epsilon$ of the
probability that~$\barsS_\Lambda$ occurs in  the product measure where the
probability of~$\sS_x=1$ is~$p_+$ if~$\bar\sigma_x=+1$ and~$p_-$
if~$\bar\sigma_x=-1$. 

Let~$(m,\theta)$ be the unique minimizer
of~$\scrG_{h,c}(m,\theta)$.  Taking expectation of
\eqref{3.33} over~$\bar\sigma$ with~$\bar\sigma_\Lambda$ fixed, using
Corollary~\ref{cor3.6} to discard the events~$|M_L/L^d-m|\ge\epsilon$
or~$|Q_L/L^d-\theta  c|\ge\epsilon$ and invoking the continuity of~$p_\pm$
in~$m$ and~$\theta$, we find out that
$P_L^{\pm,c,h}(\sS_\Lambda=\barsS_\Lambda|\sigma_\Lambda=\bar\sigma_\Lambda)$
indeed converges to
\begin{equation}
%\label{}
\prod_{x\in
\Lambda}\bigl\{p_{\bar\sigma_x}\delta_1(\barsS_x)+(1-p_{\bar\sigma_x})\delta_0(\barsS_x)\bigr\},
\end{equation}
with~$p_\pm$ evaluated at the minimizing~$(m,\theta)$.  But
for this choice Lemma~\ref{lemma3.7} guarantees that $p_\pm=q_\pm$, which
finally  proves \twoeqref{2.7}{2.8}.
\end{proofsect}

The last item to be proved is Proposition~\ref{prop2b} establishing  the
basic features of the phase diagram of the model under consideration:

\begin{proofsect}{Proof of Proposition~\ref{prop2b}} 
From Lemma~\ref{lemma3.7} we already know that the set of points~$m(h,c)=m$
for~$m\in[-\mstar,\mstar]$ is given  by the equation
\eqref{hequals}. By the fact that~$m(h,c)$ is strictly increasing  in~$h$
and that~$m(h,c)\to\pm1$ as~$h\to\pm\infty$ we thus know that
\eqref{hequals} defines a line  in the~$(h,c)$-plane. Specializing
to~$m=\pm\mstar$ gives us two curves parametrized by  functions~$c\mapsto
h_\pm(c)$ such that at~$(h,c)$ satisfying~$h_-(c)<h<h_+(c)$ the system
magnetization~$m(h,c)$ is strictly between~$-\mstar$ and~$\mstar$, i.e.,
$(h,c)$ is in the phase  separation region.  

It remains to show that the
above functions~$c\mapsto h_\pm(c)$ are strictly  monotone and negative
for~$c>0$. We will invoke the expression \eqref{hequals} which applies
because  on the above curves we have~$m(h,c)\in[-\mstar,\mstar]$. Let us
introduce new variables
\begin{equation}
%\label{}
R_+=\frac{q_+}{1-q_+}\quad\text{and}\quad R_-=\frac{q_-}{1-q_-}
\end{equation}
and, writing~$h$ in \eqref{hequals} in terms  of~$R_\pm$, let
us differentiate with respect to~$c$. (We will denote the corresponding
derivatives by  superscript prime.) Since
\eqref{2.8} gives us that~$R_-=e^{-\kappa}R_+$, we easily derive
\begin{equation}
%\label{}
2h'=\frac{R_-'}{1+R_-}-\frac{R_+'}{1+R_+}=-R_+'\frac{1-e^{-\kappa}}{(1+R_+)(1+R_-)}.
\end{equation}
Thus,~$h'$ and~$R_+'$ have opposite signs; i.e., we  want to
prove that~$R_+'>0$. But that is immediate: By the second equation in
\eqref{2.8} we  conclude that at least one of~$R_\pm'$ must be strictly
positive, and by~$R_-=e^{-\kappa}R_+$ we  find that both~$R_\pm'>0$. It
follows that~$c\mapsto h_\pm(c)$ are strictly decreasing, and
since~$h_\pm(0)=0$, they are also negative once~$c>0$.
\end{proofsect}

\section*{Acknowledgments}
\noindent The research of K.S.A. was
supported by the NSF under the grants DMS-0103790 and DMS-0405915.  The
research of M.B. and L.C.~was supported by the NSF grant~DMS-0306167.


\smallskip
\begin{thebibliography}{aaa}

\bibitem{ABC2} K.S.~Alexander, M.~Biskup and L.~Chayes,
\textit{Colligative properties of solutions~II.~Vanishing concentrations},
next paper.

\bibitem{ACC} K.~Alexander, J.T.~Chayes and L.~Chayes, \textit{The  Wulff
construction and asymptotics of the finite cluster distribution for
two-dimensional  Bernoulli percolation},
\textrm{Commun. Math. Phys.} \textbf{131} (1990) 1--51.

\bibitem{BCG} Ph.~Blanchard, L.~Chayes and D.~Gandolfo, \textit{The  random
cluster representation for the infinite-spin Ising model: Application to QCD
pure gauge theory}, Nucl. Phys.~B~[FS]~\textbf{588} (2000) 229--252.

\bibitem{bigBCK} M.~Biskup, L.~Chayes and R.~Koteck\'y,
\textit{Critical region for droplet formation  in the two-dimensional  Ising
model},  Commun. Math. Phys. \textbf{242} (2003), no. 1-2, 137--183.

\bibitem{BCK-comment} M.~Biskup, L.~Chayes, and R.~Koteck\'y,
\textit{Comment on: ``Theory of the evaporation/condensation transition of
equilibrium droplets in finite  volumes''}, Physica A
\textbf{327} (2003) 589-592.

\bibitem{Bodineau} T.~Bodineau, \textit{The Wulff construction in  three and
more dimensions},
\textrm{Commun. Math. Phys.} \textbf{207} (1999) 197--229.

\bibitem{Bodineau2} T.~Bodineau, 
\textit{Slab percolation for the Ising model}, math.PR/0309300.

\bibitem{BIV}  T.~Bodineau, D.~Ioffe and Y.~Velenik,
\textit{Rigorous probabilistic analysis of equilibrium crystal  shapes},
\textrm{J.~Math. Phys.}
\textbf{41} (2000) 1033--1098.

\bibitem{Cerf} R.~Cerf, \textit{Large deviations for three  dimensional
supercritical percolation}, \textrm{Ast\'erisque} \textbf{267} (2000)
vi+177.

\bibitem{Cerf-Pisztora} R.~Cerf and A.~Pisztora, \textit{On the Wulff
crystal in the Ising model}, \textrm{Ann. Probab.} \textbf{28} (2000)
947--1017.

\bibitem{Cottrell} A.H.~Cottrell, \textit{Theoretical Structural
Metallurgy}, St.~Martin's Press, New York, 1955.

\bibitem{Curie} P.~Curie, \textit{Sur la formation des cristaux et  sur les
constantes capillaires de leurs diff\'erentes faces}, Bull. Soc.
Fr.~Mineral. \textbf{8}  (1885) 145; Reprinted in
\textit{\OE uvres de Pierre Curie}, Gauthier-Villars, Paris, 1908,
pp.~153--157.

\bibitem{Dembo-Zeitouni} A.~Dembo and O.~Zeitouni, \textit{Large  Deviations
Techniques and Applications} (Springer Verlag, Inc., New York,~1998).

\bibitem{DKS} R.L.~Dobrushin, R.~Koteck\'y and S.B.~Shlosman,
\textit{Wulff construction. A global shape from local interaction}, Amer.
Math. Soc., Providence, RI, 1992.

\bibitem{DS} R.L.~Dobrushin and S.B.~Shlosman, In:
\textit{Probability contributions to statistical mechanics}, pp.~91-219,
Amer. Math. Soc., Providence, RI, 1994.

\bibitem{Gibbs} J.W. Gibbs, \textit{On the equilibrium of  heterogeneous
substances} (1876),  In:
\textit{Collected Works},  vol.~1., Longmans, Green and Co., 1928.

\bibitem{Golden} K.M.~Golden, \textit{Critical behavior of transport in sea
ice}, \textrm{Physica B} \textbf{338} (2003)~274--283.

\bibitem{GoldenAckleyLytle} K.M.~Golden, S.F.~Ackley and V.I.~Lytle,
\textit{The percolation phase transition in sea ice}, \textrm{Science}
\textbf{282} (1998) 2238--2241.

\bibitem{GS1} P.E.~Greenwood and J.~Sun,
\textit{Equivalences of the large deviation principle for Gibbs  measures
and critical balance in the Ising model}, J.~Statist. Phys. \textbf{86}
(1997),  no.~1-2, 149--164.

\bibitem{GS2} P.E.~Greenwood and J.~Sun,
\textit{On criticality for competing influences of boundary and  external
field in the Ising model}, J.~Statist. Phys. \textbf{92} (1998), no.~1-2,
35--45.

\bibitem{denHollander} F.~den~Hollander, \textit{Large Deviations},  Fields
Institute Monographs, vol~14, American Mathematical Society, Providence, RI,
2000.

\bibitem{Bob+Tim} D.~Ioffe and R.H.~Schonmann,
\textit{Dobrushin-Koteck\'y-Shlosman theorem up to the critical
temperature}, \textrm{Commun. Math. Phys.} \textbf{199}  (1998) 117--167.

\bibitem{Kotecky-Medved} R.~Koteck\'y and I.~Medved'\!,
\textit{Finite-size scaling for the 2D Ising model with minus boundary
conditions}, J.~Statist. Phys.
\textbf{104} (2001), no.~5/6, 905--943.

\bibitem{Landau-Lifshitz} L.D.~Landau and E.M.~Lifshitz,
\textit{Statistical Physics}, Course of Theoretical Physics, vol.~5,
Pergamon Press, New York, 1977.

\bibitem{LY} T.D.~Lee and C.N.~Yang, \textit{Statistical theory of
equations of state and phase transitions: II. Lattice gas and Ising model},
Phys. Rev. \textbf{87}  (1952)  410--419.

\bibitem{MatsumotoSaitoOhmine} M.~Matsumoto, S.~Saito and I.~Ohmine,
\textit{Molecular dynamics simulation of the ice nucleation and proces
leading to water freezing}, Nature \textbf{416} (2002) 409--413.

\bibitem{Moore} W.J.~Moore, \textit{Physical Chemistry} (4th  Edition),
Prentice Hall, Inc., Englewood Cliffs, NJ 1972.

\bibitem{Pfister} C.-E. Pfister, \textit{Large deviations and phase
separation in the two-dimensional Ising model}, \textrm{Helv. Phys. Acta}
\textbf{64} (1991) 953--1054.

\bibitem{Pf-Velenik} C.-E.~Pfister and Y.~Velenik, \textit{Large  deviations
and continuum limit in the~$2$D Ising model}, \textrm{Probab. Theory Rel.
Fields}
\textbf{109} (1997) 435--506.

\bibitem{SS2} R.H.~Schonmann and S.B.~Shlosman,
\textit{Constrained variational problem with applications to the  Ising
model}, J.~Statist. Phys.
\textbf{83}  (1996),  no. 5-6, 867--905.

\bibitem{Sinai} Ya.G.~Sina\u\i, \textit{Theory of Phase Transitions:
Rigorous Results}, International Series in Natural Philosophy, vol.~108,
Pergamon Press,  Oxford-Elmsford, N.Y., 1982.

\bibitem{Wortis} C.~Rottman and M.~Wortis, \textit{Statistical  mechanics of
equilibrium crystal shapes: Interfacial phase diagrams and phase
transitions}, Phys. Rep.
\textbf{103} (1984) 59--79.

\bibitem{Wulff} G.~Wulff, \textit{Zur Frage des Geschwindigkeit des
Wachsturms und der Aufl\"osung der Krystallflachen}, \textrm{Z.~Krystallog.
Mineral.}
\textbf{34} (1901) 449--530.

\end{thebibliography}
\end{document}